\documentclass[%
 aip,
 amsmath,amssymb,
reprint,%
]{revtex4-2}
\usepackage{graphicx}  
\usepackage{dcolumn}   
\usepackage{bm}        
\usepackage[colorlinks]{hyperref}
\usepackage[all]{hypcap}
\usepackage{color}
\usepackage{amsmath}
\usepackage{physics}
\usepackage{lipsum}
\usepackage{pdfcomment}

\usepackage{color}

\pdfcommentsetup{draft}

\newcommand{\cqt}{Centre for Quantum Technologies, National University of Singapore, 3 Science Drive 2, Singapore 117543, Singapore}
\newcommand{\ntu}{Division of Physics and Applied Physics, School of Physical and Mathematical Sciences, Nanyang Technological University, 21 Nanyang Link, Singapore 637371, Singapore}
\newcommand{\aqs}{Aqsolotl Pte.~Ltd.~160 Robinson Road, \#14-04 SBF Center,  Singapore 068914, Singapore}

\begin{document}

\preprint{APS/123-QED}

\title{Characterizing Noise in Controlling Superconducting Qubits}

\author{Yuanzheng Paul \surname{Tan}}
\affiliation{\ntu{}}

\author{Yung Szen \surname{Yap}}
\email[The author to whom correspondence may be addressed: ]{yungszen@utm.my}
\affiliation{Faculty of Science and Centre for Sustainable Nanomaterials (CSNano), Universiti Teknologi Malaysia, 81310 UTM Johor Bahru, Johor, Malaysia}
\affiliation{\cqt{}}

\author{Long Hoang \surname{Nguyen}}
\affiliation{\cqt{}}

\author{Rangga P. \surname{Budoyo}}
\affiliation{\cqt{}}

\author{Patrick \surname{Bore}}
\affiliation{\aqs{}}

\author{Kun Hee \surname{Park}} 
\affiliation{\cqt{}}

\author{Christoph \surname{Hufnagel}}
\affiliation{\cqt{}}

\author{Rainer \surname{Dumke}}
\affiliation{\cqt{}}
\affiliation{\ntu{}}

\begin{abstract}
Meaningful quantum computing is currently bottlenecked by the error rates of current generation Noisy Intermediate Scale Quantum (NISQ) devices. 
To improve the fidelity of the quantum logic gates, it is essential to recognize the contributions of various sources of errors, including background noise.
In this work, we investigate the effects of noise when applied to superconducting qubit control pulses to observe the dependency of the gate fidelity with the signal-to-noise ratio (SNR).
We propose a model on how the noise of the control electronics interacts with the qubit system and demonstrate a method for characterizing the noise environment of the qubit control.

\begin{description}
\item[Keywords]
Superconducting qubit; Thermal Noise; Signal-To-Noise Ratio; Fidelity;
\end{description}
\end{abstract}

\maketitle



\section{Introduction}

Computations on quantum hardware offers a promising speed-up for certain tasks as compared to classical system. 
However, the Noisy Intermediate Scale Quantum (NISQ) devices that are currently used for quantum computation suffer from high error rates for meaningful computation \cite{Vandersypen2001,PhysRevA.94.012314, dalton2024quantifying}.
In order to bridge the gap between NISQ devices of today and the fault-tolerant devices of the future, it is essential to break down and examine the current causes of the qubit errors and their extent.

At its core, quantum computation involves the manipulation of qubits using quantum logic gates~\cite{Nielsen_Chuang_2010}.
For superconducting qubits, any single-qubit quantum logic gate can be realized with two microwave pulses and phase shifts on these pulses~\cite{krantz2019engguide}.
With regard to the accuracy of the computation, we examine the interaction between superconducting qubits~\cite{Nakamura_1999} and the microwave control system.

The quality of the qubit is characterized by its (de)coherence times: energy relaxation time, $T_1$, and dephasing time, $T_2$,~\cite{burnett2019decoherence} where longer coherence times are able to support a longer/deeper circuit.
On the other hand, the quality of the quantum gates are characterized by their error rates.
%
%
%
Also known as gate fidelity, this figure of merit is determined not only by the microwave control system but is also influenced by the qubit coherence times. 
As such, to achieve fault-tolerant quantum computation (FTQC) by implementing quantum error correction, one would need good gate fidelities, long coherence times and more physical qubits.
Presently, a typical quantum processor has coherence times in the order of tens to hundreds of microseconds~\cite{Wang2022qblifetime} with single-qubit and two-qubit fidelities in the range of 99.9\% and 99.8\%~\cite{marxer2023long},  respectively.

Quantum gate errors are of particular interest in this work.
Some of them arise from the signal imperfection that are applied to the qubit, such as:
\begin{itemize}
\item inaccurate pulse length or shape~\cite{yapBB1}; 
\item off-resonance and spectral purity (phase noise/jitter)~\cite{berger2015geometric, PhysRevApplied.3.044009, viotti2023geometric, IEEEPhaseNoise};
\item crosstalk between microwave channels~\cite{PhysRevApplied.18.024068};
\item noise and spurious signals; and 
\item drift/performance stability.
\end{itemize}

In this paper, we investigated the qubit and pulse carrier frequency drifts and the impact of signal-to-noise ratio (SNR) towards gate fidelities.
We proposed a model to understand how noise affects the qubit gate fidelity.
Based on this model, we explained how we measured these sources of errors on a superconducting qubit with finite coherence times.
We performed a Ramsey experiment to measure the frequency fluctuation of the qubit and extend the model to simulate its effect on the gate fidelity. 
Finally, we varied the SNR by varying the noise power and measured the resulting gate fidelities.
From the measured fidelities as a function of SNR, we hope to better understand the importance of these sources of errors on gate fidelity and estimate the required SNRs to achieve 0.1\% and 0.01\% error rates.

\section{Model}

For this work, we used a 3D qubit in a cavity with coherence times of $T_1 = (8.66 \pm 0.49)~\mu$s and $T_2 = (9.08 \pm 0.74)~\mu$s, controlled using microwave pulses. The interaction between the microwave field and the superconducting qubit yields the following Hamiltonian~\cite{krantz2019engguide}:

\begin{equation}
    \mathcal{H}_d = -\Omega_R V(t) \begin{bmatrix}
        0 & e^{i(\delta\omega t + \phi)} \\
        e^{-i(\delta\omega t + \phi)} & 0
    \end{bmatrix}
    \label{eqn:drive-with-detuning}
\end{equation}

\noindent where $\Omega_R$ is the Rabi frequency of the qubit, $V(t)$ is the time-dependent amplitude of the pulse, $\delta\omega$ is the detuning between the frequency of the control pulse and the transition frequency of the qubit and $\phi$ is the phase of the the microwave. If the frequency of the pulse is on resonance with the qubit, the Hamiltonian becomes

\begin{equation}
    \mathcal{H}_d =  -\Omega_R V(t) (\cos{\phi}\sigma_x + \sin{\phi}\sigma_y)
    \label{eqn:drive}
\end{equation}

\noindent where $\sigma_x$ and $\sigma_y$ are the Pauli X and Y operators, respectively. 
By choosing the phase of the pulse, $\phi$, we select the axis of the rotation of the qubit about the XY-plane in the Bloch sphere. 
The amplitude $V(t)$ modulates the speed of the rotation and the pulse duration controls the evolution time. 
In the following sections, we discuss the source of each error and how it affects the qubit control.


\section{Errors in Controlling Superconducting Qubits}

To measure the error rate of a single-qubit gate fidelity, we performed the randomized benchmarking protocol on the 3D qubit. 
The protocol consists of a number of random Clifford gates, followed by a gate to return the qubit to the ground state. 
With errors, it does not return to the ideal ground state and from the data fit,  one can determine the gate fidelity.

For the system described in the previous section, we investigated the errors due to the (short) coherence times and microwave control system.
We obtained an experimental single-qubit gate fidelity of $\mathcal{F}^{\mathrm{exp}}_0 = (99.833 \pm 0.014)\%$ (without any additive noise). 
Simulation of the same pulse sequence using QuTiP~\cite{JOHANSSON20131234qutip} with only the mean decoherence times of the qubit (i.e.~without error from the control system) yielded a gate fidelity of $\mathcal{F}^{\mathrm{sim}}_0 =(99.849 \pm 0.001)\%$. 
This value tells us that the short coherence times contributed an error rate of $\varepsilon_{\textrm{cor}} = (0.151 \pm 0.001) \%$.
Furthermore, it is important to note that the difference between the experimental and simulation gate fidelity is $\varepsilon_{\mathrm{others}} = (0.016 \pm 0.014)\%$, which gives us the approximate error rate from \emph{all other sources} including our control system~\cite{park2022icarusq}.
The results are summarized in Table~\ref{tab:noNoise}.

\begin{table}[!h]
\caption{Randomized Benchmarking (RB) results. The simulation and experimental fidelities without any additive noise were obtained to determine the errors due to short coherence times, $\varepsilon_{\mathrm{cor}}$ and  microwave control system, $\varepsilon_{\mathrm{others}}$.}
 \label{tab:noNoise}
\begin{tabular}{|l|c|c|}
\hline

\textbf{Description}                                      																					& \textbf{Symbol} 						& \textbf{Fidelity (\%)}	\\ \hline
\begin{tabular}[c]{@{}l@{}}Simulated RB \\ (with coherence times)\end{tabular}    													&  $\mathcal{F}^{\mathrm{sim}}_0$             &  $99.849 \pm 0.001$	\\ \hline
\begin{tabular}[c]{@{}l@{}}Error due to coherence times \\ ($1-\mathcal{F}^{\mathrm{sim}}_0$) \end{tabular}   							&  $\varepsilon_{\mathrm{cor}}$              	& $0.151 \pm 0.001$       \\\hline
\begin{tabular}[c]{@{}l@{}}Experimental RB\\ (only intrinsic noise \\ i.e.~without additive noise)\end{tabular} 													&  $\mathcal{F}^{\mathrm{exp}}_0$ 		& $99.833 \pm 0.014$	\\ \hline
\begin{tabular}[c]{@{}l@{}}Errors from other sources \\ ($\mathcal{F}^{\mathrm{sim}}_0 - \mathcal{F}^{\mathrm{exp}}_0$)\end{tabular}		&  $\varepsilon_{\mathrm{others}}$		& $0.016 \pm 0.014$		\\ \hline
\end{tabular}
\end{table}

To further investigate the cause of these errors, we performed the following experiments.
First, we measured the stability of the qubit frequency over an extended period.
Then, we assessed the stability of the control system by measuring the pulse carrier frequency. 
Finally, we added noise to the control pulses to find the correlation between the gate fidelity and the signal-to-noise ratio (SNR).

\subsection{Frequency Stability}
A mismatch between the control pulse frequency and the qubit frequency causes off-resonance error in controlling the qubit. 
To avoid this, these frequencies are calibrated before an experiment, but will drift over time. 
This drift occurs for both the qubit and the control system, thus emphasizing the need to determine the error contributions as well as the necessity to perform periodic calibrations~\cite{burnett2019decoherence}.

In the superconducting qubit model, fluctuations and uncertainties of the qubit frequency may originate from microscopic effects such as defects in the junctions~\cite{burnett2019decoherence}.
In order to properly control the qubit, the control system frequency drift must be less than that of the qubit frequency drift. 
One of the cause of the control system frequency drift is the temperature fluctuations of the electronics, especially within the clocking subsystem, which includes components responsible for generating, distributing, conditioning, and controlling the clock signals.

\begin{figure}[!h]
\centering
    \includegraphics[width=\linewidth]{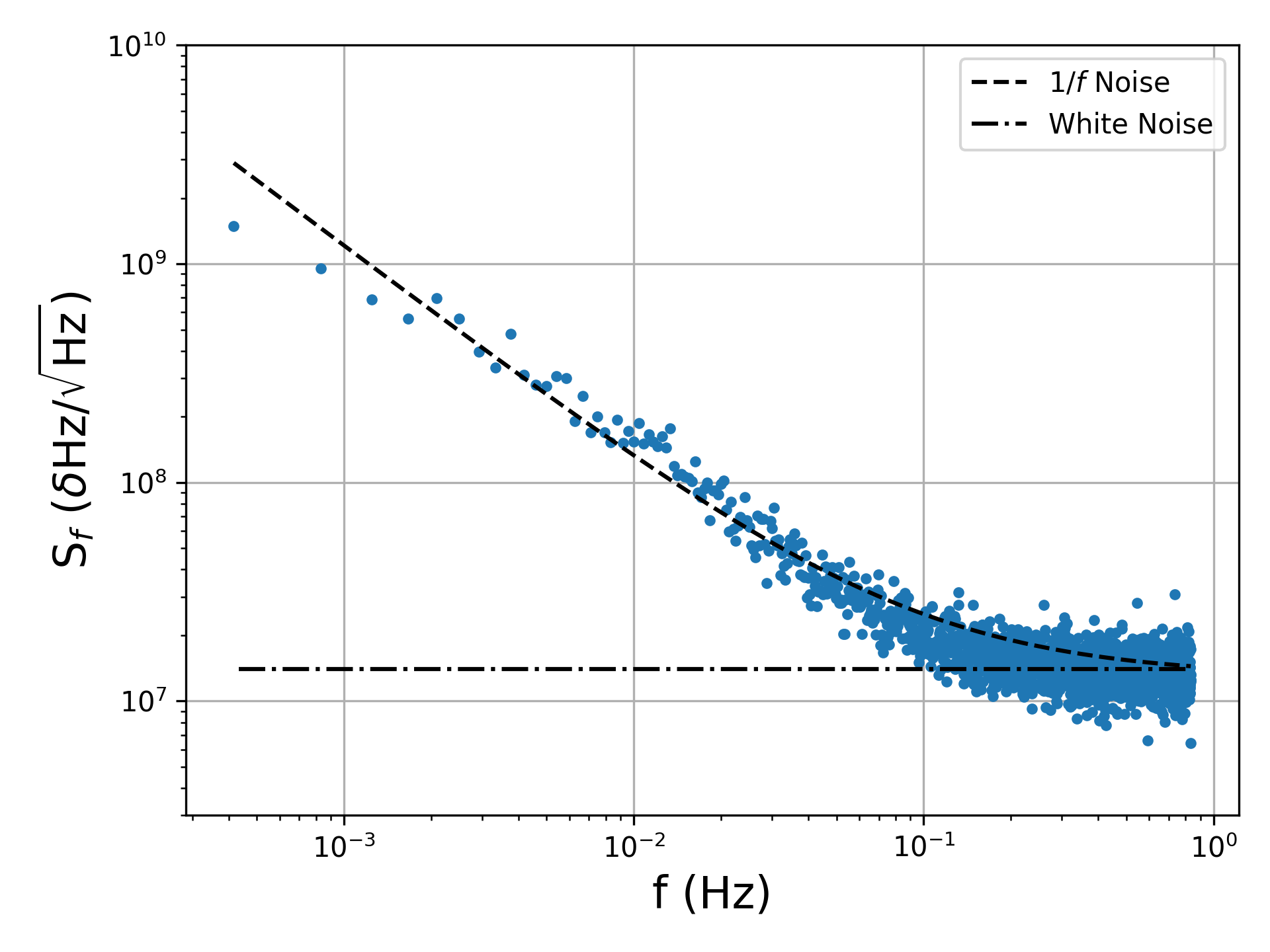}
    \caption{Frequency stability of the qubit over 20 hours indicating that the pink noise, 1/f, dominating over the white noise. Throughout the measurement, the qubit frequency drifted around $\pm 5$~kHz, corresponding to one-sigma deviation.}
    \label{fig:ramsey}
\end{figure}

\begin{figure}[!h]
 \centering
     \includegraphics[width=\linewidth]{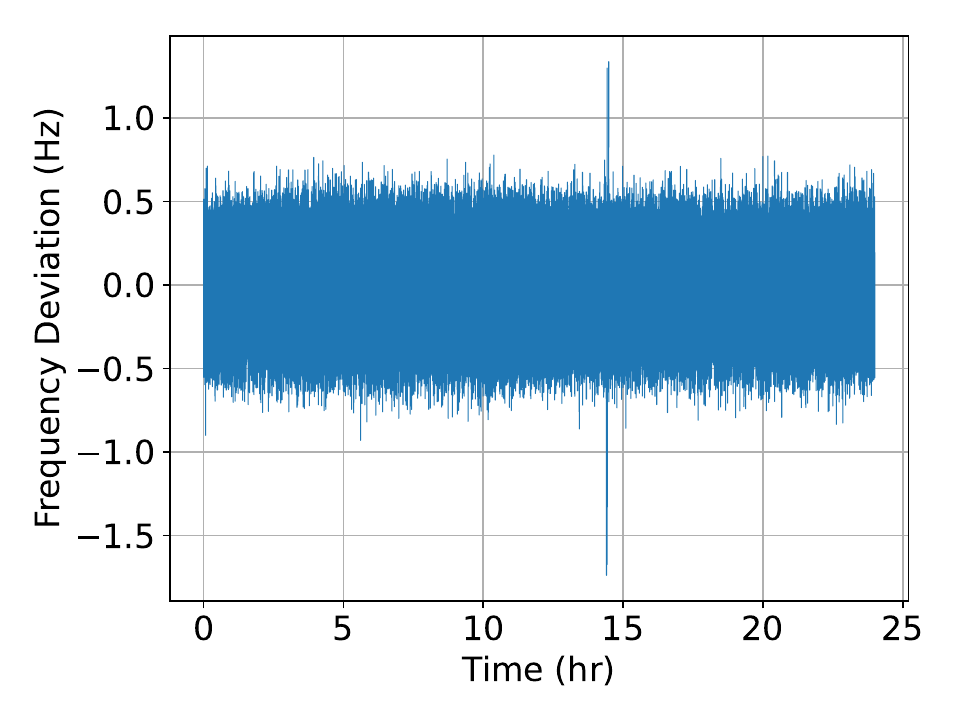}
     \caption{Frequency drift of the control system in 24 hours. The control system frequency drift range around $\pm 0.2$~Hz over one-sigma deviation, which is negligible as compared to qubit frequency drift.}
     \label{fig:drift}
 \end{figure}

To measure the frequency stability of the qubit, we ran the Ramsey experiment on the qubit for over 20 hours, presented in Figure~\ref{fig:ramsey}.
Throughout the measurement, the qubit frequency drifted around $\pm 5$~kHz, corresponding to one-sigma deviation.
The spectral density shows that the noise of the qubit frequency was found to consist of white and pink noises, with pink noise being dominant.
Pink noise has been attributed to the microscopic effects of the qubit~\cite{burnett2019decoherence} while white noise may be attributed to the thermal noise of control setup electronics.
Similarly, we measured the frequency stability of the control system for 24 hours (Figure~\ref{fig:drift}). 
We found that the controller frequency drifts around $\pm 0.2$~Hz over one-sigma deviation, which is insignificant compared to the qubit frequency drift. 

\subsection{Additive Noise}

The pulses used to control the qubit intrinsically contained noise and other artifacts, which resulted in a finite SNR.
In this experiment, we added noise to the control pulses to vary its SNR and examined its corresponding gate fidelities.

\begin{figure}[!tb]
    \centering
    \includegraphics[width=\linewidth]{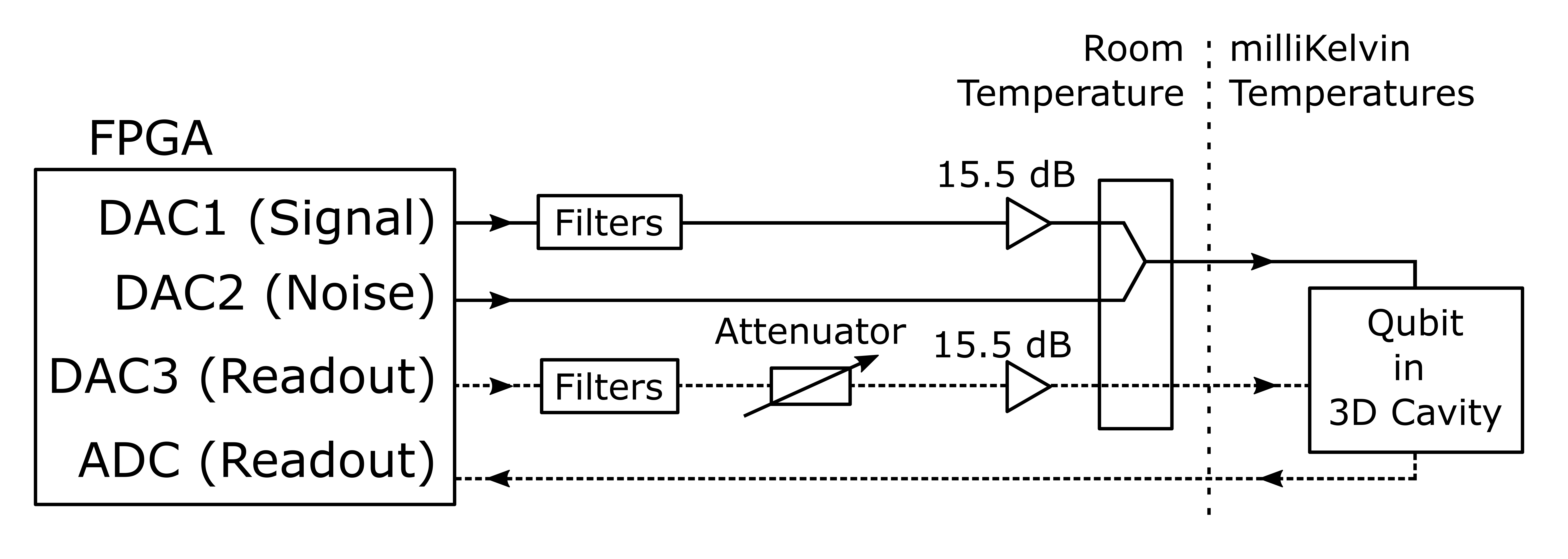}
    \caption{Experimental setup to investigate effects of white noise towards error rates.  The DAC1 output channel was used to generate the control pulse at 4.564~GHz and -14.77~dBm.  It was filtered, amplified and combined with noise generated from DAC2 output channel.  The control pulse with additive noise was directed to the superconducting qubit cooled to about 10~mK inside a dilution refrigerator.  The microwave components in the dilution refrigerator are not shown for clarity. A readout pulse (at 5.1211~GHz) from the DAC3 output channel was sent to a resonator coupled to the superconducting qubit and by evaluating the returned pulsed to the ADC input channel, the state of the qubit was determined. For this experiment, the power of the additive noise was varied to find the correlation between the error rate and the signal-to-noise ratio of the control pulse.}
    \label{fig:circuit}
\end{figure}

\begin{figure*}[!bt]
    \centering
    \includegraphics[width=\textwidth]{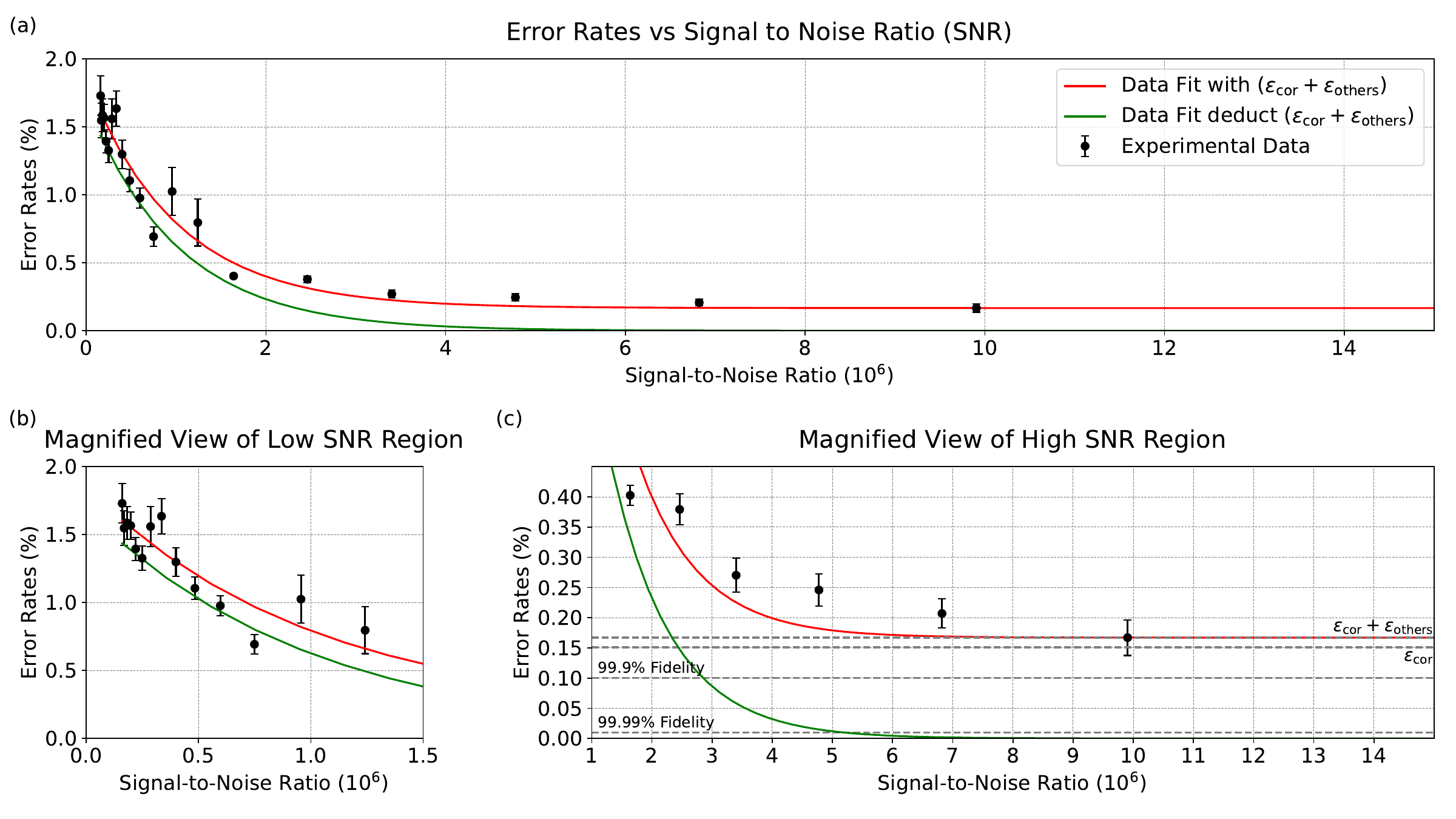}
    \caption{Single-qubit gate fidelities with decreasing signal-to-noise ratio (SNR). 
The SNR was estimated based on the signal power of -14.77 dBm plus a gain of 15.5 dB, over the root-mean-square noise power within 100~MHz bandwidth.
The full SNR range is plotted in (a) whereas two magnified views are plotted in (b) and (c) for the low and high SNR regions, respectively.
The solid red line data fit was plotted using $y=A\cdot{}\exp(-x\cdot{}b) +\varepsilon_{\mathrm{cor}}  + \varepsilon_{\mathrm{others}}$. 
The $\varepsilon_{\mathrm{cor}}$ is the error rate caused by the coherence time of the qubit and $\varepsilon_{\mathrm{others}}$ is the approximate error rate from all other
sources including our control system.    
The least square method obtained the following fitting constants: $A=1.682, b = 0.9898$.  
In (c), the green dashed line is the  solid red line data fit without $\varepsilon_{\mathrm{cor}}+\varepsilon_{\mathrm{others}}$ and the SNRs required for 99.9\% and 99.99\% fidelities are approximately $2.9\times10^6$ and $5.2\times10^6$, respectively.} 
    \label{fig:fidelity-snr}
\end{figure*}

This experiment was done using two separate digital-to-analog converter (DAC) channels: one DAC for generating the control pulses and the other DAC for generating the white noise waveform.  
The pulse and noise were combined using a microwave combiner (see Figure~\ref{fig:circuit}) and randomized benchmarking was carried out to determine the gate fidelity.
By increasing the amplitude of the additive noise, we reduced the SNR of the control pulse and measured its respective fidelity.
This allows us to examine the relationship between the error rates at different SNRs (see Fig.~\ref{fig:fidelity-snr}).

From the experiment, the error rates were found to decrease exponentially with SNR. 
The additive noise caused the gate operation to randomly excite the qubit between the ground and first excited states and deviating away from its targeted operation.
Furthermore, the additive noise also caused the qubit to excite to higher energy states which is not recoverable in the randomized benchmarking sequence, thus resulting in low fidelities.
In the high SNR region, the error rates reached a minimum value without any additive noise i.e.~this error rate is the sum of errors from the qubit decoherence, $\varepsilon_{\mathrm{cor}}$ and other intrinsic error, $\varepsilon_{\mathrm{others}}$ of the entire system.

For an ideal qubit without decoherence, the required SNR to achieve error rates of 0.1\% and 0.01\% are approximately $2.9\times{}10^6$ and $5.2\times{}10^6$, respectively.
In reality, after achieving these SNR values, we expect some other microwave signal imperfections to become dominant from the control system.


\section{Conclusion}
In this work, we investigate how noise from: (i) the frequency stability of the qubit and microwave control system, (ii) the thermal noise of the control system and (iii) the (de)coherence times of the qubit, contribute to the gate fidelities.
From the frequency stability measurements, we found that our control system has lower drift compared to the qubit frequency.
This indicates that calibration to measure the qubit frequency needs to be done periodically.
The same experiment spanning over 60 hours was done by Ref.~\cite{burnett2019decoherence} and our work here is an extension to that, where in addition of the qubit frequency, we also conclusively determined the frequency stability of the control system.
Furthermore, we also demonstrated that the gate fidelity is affected by the SNR of the control pulses and the coherence times of the qubit. 
To achieve gate fidelities of 99.9\% and higher, we would require qubits with longer T$_1$ and T$_2$ coherence times.
Experimentally, we were unable to resolve the intrinsic error, $\varepsilon_{\mathrm{others}} = (0.016 \pm 0.014)\%$, especially since the uncertainty is comparable to the measured error rate.
Nevertheless, it would be interesting to investigate other possible sources of errors including microwave artifacts and its effects towards gate fidelities.

\begin{acknowledgments}
This work was supported by the National Research Fund (NRF) Quantum Engineering Programme (QEP)  W21Qpd0211 Advanced Quantum Processor Platform.
\end{acknowledgments}

\section*{Author Declaration}

\subsection*{Conflict of Interest}
The authors have no conflicts to disclose.

%

\bibliography{main}

\end{document}